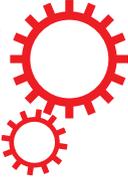

**OPEN** # Superconducting/magnetic Three-state Nanodevice for Memory and Reading Applications

J. del Valle[1], A. Gomez[1,†], E. M. Gonzalez[1,2], M. R. Osorio[2], D. Granados[2] & J. L. Vicent[1,2]

We present a simple nanodevice that can operate in two modes: i) non-volatile three-state memory and ii) reading device. The nanodevice can retain three well defined states −1, 0 and +1 and can operate in a second mode as a sensor for external magnetic fields. The nanodevice is fabricated with an array of ordered triangular-shaped nanomagnets embedded in a superconducting thin film gown on Si substrates. The device runs based on the combination of superconducting vortex ratchet effect (superconducting film) with the out of plane magnetization (nanomagnets). The input signals are ac currents and the output signal are dc voltages. The memory mode is realized without applying a magnetic field and the nanomagnet stray magnetic fields govern the effect. In the sensor mode an external magnetic field is applied. The main characteristic of this mode is that the output signal is null for a precise value of the external magnetic field that only depends on the fabrication characteristics of the nanodevice.

Nowadays, one of the main topics in nanotechnology is the fabrication and performance of robust and versatile nanodevices with different functionalities. We have fabricated a non-volatile memory and reading device, which is designed taking into account the interplay between ratchet of superconducting vortices and stray magnetic fields generated by nanomagnets with out of plane magnetization. This superconducting/magnetic device benefits from the noteworthy properties of ratchet effect. Worth to notice that proposals for different devices can be found based on the interplay between ratchet effect and vortices[1–2] as well as colloids[3]. Ratchet is a very general effect, which is in the core of many interesting phenomena and applications[4]. Ratchet effect occurs when, driven by zero-average force, particles are moving on asymmetric potentials. Different origins can provide the driving forces; for instance, via an external time-dependent modulation (ac injected current in our case)[5], or by energy input from a non-equilibrium source such as a chemical reaction (motor proteins)[6]. Concerning the asymmetric potentials, ratchet effect can be classified in rocking ratchets when the potential is time independent (triangle-shaped nanomagnets, in the present case) or flashing ratchets when the potential is time dependent, for instance when the potential is turned on and off, as happens in organic electronic based ratchet[7]. In our rocking ratchet device the particles are superconducting vortices present in superconducting Nb thin film, the asymmetric potentials are provided by array of nanotriangles fabricated with Co/Pd multilayers and the input signal is ac current. As is shown the key element for this nanodevice is the out of plane remanent magnetic state of the Pd/Co multilayers[8], (see supplementary information for the magnetic characterization). In summary, we have built with these ingredients a nanodevice that acts as a non-volatile memory with three resilient states and moreover, when a magnetic field is applied, the nanodevice operates as magnetic field sensor.

The paper is organized as follows: First we briefly describe the nanodevice, second we show the two working modes, then we depict the memory mode, next we show how the sensor mode works and finally we explain the effects which govern the device. A summary closes the paper.

[1]Departamento Fisica de Materiales, Facultad de CC. Fisicas, Universidad Complutense, 28040 Madrid (Spain). [2]IMDEA-Nanociencia, Cantoblanco, 28049 Madrid (Spain). †Present address: Centro de Astrobiología (CSIC-INTA), Torrejón de Ardoz, 28850 Madrid (Spain). Correspondence and requests for materials should be addressed to J.L.V. (email: jlvicent@ucm.es or joseluis.vicent@imdea.org)

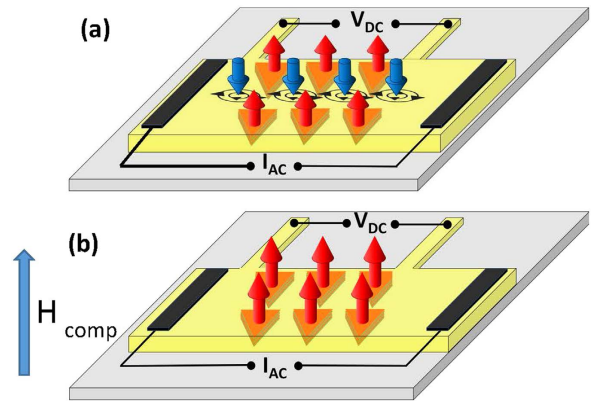

**Figure 1. Sketch of the nanodevice.** Co/Pd multilayer triangular–shaped nanomagnets (orange). Nb thin film (yellow), Si substrate (green) (sketch not to scale). Out of plane magnetization (red arrows). Superconducting vortices (blue arrows). H is the applied magnetic field. (**a**) H = 0, (**b**) H ≠ 0, in the case H = H $_{comp}$. H $_{comp}$ being the magnetic field needed to annihilate the interstitial vortices.

## Results

The array of magnetic nanotriangles on Si substrate is grown by sputtering and electron beam lithography. The array periodicity, size and thickness of the nanotriangles are similar to the array of Ni nanotriangles reported in a previous work[5]. A 100 nm thick Nb superconducting film is sputtered on top of the array. Finally, we define, by conventional lithography and reactive etching techniques a bridge[5] which allows injecting ac currents and measuring dc voltages in the device. In summary the nanodevice is a Si substrate with an array of triangular nanomagnets embedded in a superconducting film (see methods for details).

**Device working modes.** Superconductivity and magnetism are two cooperative effects which compete with each other. Recently, this antagonist behavior has been turned around, since magnetic nanostructures can enhance and control superconducting properties[9–12]. In this device, when triangles are magnetized in the upward direction, the magnetic stray field in between triangles will point downwards, and superconductivity weakens. Applying external magnetic field this effect can be compensated, hence inducing superconductivity. The nanodevice is based on this superconductivity/magnetism interplay. Two situations can be studied i) without applied magnetic field (first mode) and ii) with applied magnetic field (second mode).

In the first mode, remanent out of plane magnetization will generate stray fields threading the sample perpendicularly to the film plane. Fig. 1(a) depicts a sketch of this situation. Negative stray fields lead to vortices in between the nanotriangles; so that, particles (vortices) are present regardless of an external field is absent. In our ratchet experiments the input signal is ac current and the output signal is dc voltage. The ac current density ($J_{ac}$) produces an ac Lorentz force ($F_L$) on the vortices that is given by $F_L = J_{ac} \times n \Phi_0$; $n$ being unitary vector in the direction of the vortex and $\Phi_0 = 20.7\,G\mu m^2$. Taking into account the expression[13] for the electric field $E = B \times v$, $B$ being the magnetic induction and $v$ being the vortex-lattice velocity; the dc voltage drop ($V_{dc}$), measured along the direction of the injected current, is proportional to the vortex-lattice velocity in the direction of the ac driving force. For ac current input ($J_{ac}$) the output is a nonzero dc voltage $V_{dc}$. So that, a net flow of vortices arises from the ac driving force. We have to point out that this ratchet effect is adiabatic[5], we have measured injecting ac currents between 10 and 100 kHz, and the output dc voltages do not change.

In the second mode a magnetic field is applied. There are two possible experimental situations which are related to the direction of the applied field. Magnetic fields applied parallel and in the same direction to the triangle magnetization and magnetic fields applied parallel and in the opposite direction to the triangle magnetization. In both situations vortices will be generated and applying ac current ratchet effect will arise. We have to note that the interesting situation for our device is when the applied magnetic field compensates the stray fields, and therefore there are not interstitial vortices, Fig. 1(b) depicts this case.

**Three-state memory mode (H = 0).** Figure 2(a) shows the experimental ratchet effect without applying magnetic fields. Three different outcomes emerge i) positive ratchet effect, ii) zero ratchet effect, and iii) negative ratchet effect. The three findings are governed by the magnetization states of the nanotriangles. Since the nanotriangles show out of plane magnetic anisotropy, the nanodevice can be set up in three different magnetic states: i) magnetization upward ($+M_z$), ii) demagnetized state ($M_z = 0$) and iii) magnetization downward ($-M_z$). The remanent up and down states are easily obtained applying magnetic fields in the perpendicular direction to the sample plane upward ($+H_z$) and downward ($-H_z$) respectively. The demagnetized state is reached in the usual way by minor hysteresis loops decreasing the strength of the applied magnetic field up to zero. The final magnetic state of the device is checked using a SQUID magnetometer. If $M_z = 0$ (demagnetized state) we get $V_{dc} = 0$, if $M_z \neq 0$, then non-zero ratchet signal is measured which polarity changes with the reversal of $M_z$.

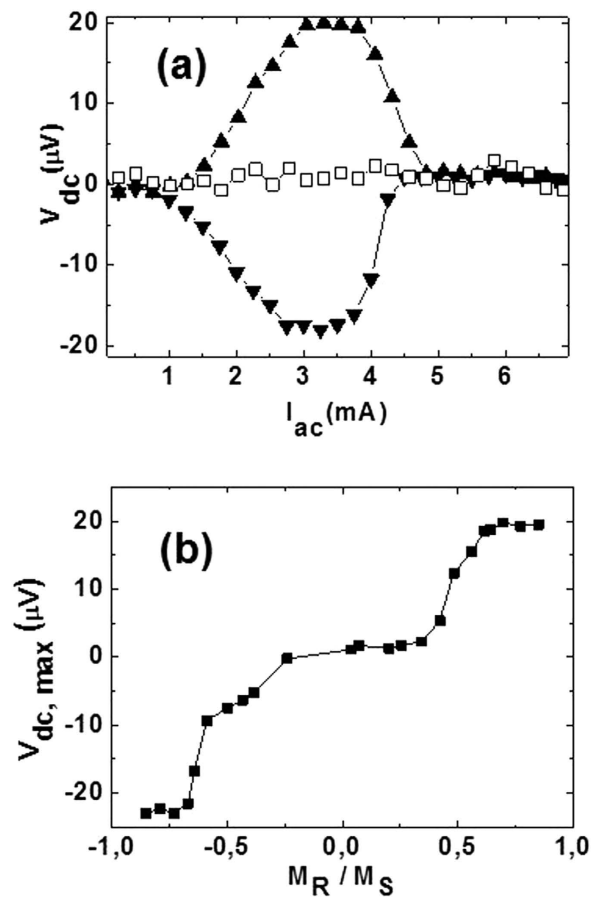

**Figure 2.** Ratchet effect without applied magnetic field: (**a**) dc voltages (Y-axis) as a function of the applied ac currents (X-axis) for $M = M_Z$ (upward triangles), $M = -M_Z$ (downward triangles) and $M = 0$ (hollow squares); $M_Z/M_S = 0.65$ ($M_s$ being the magnetization at saturation); (**b**) Maximum dc voltage (Y-axis) as a function of the normalized remanent magnetization (X-axis) for $T = 8.45$ K. ac current frequency is 10 kHz.

In Fig. 2(b) we have plotted the maximum values of the ratchet signal vs. the remanent magnetization of the nanomagnets. First of all, using a SQUID magnetometer different remanent magnetic states are measured and selected, for example: $M_z(R1)$, $M_z(R2)$, $M_z(R3)$, etc. Both branches (positive and negative) are obtained in the same way. In the following we describe the positive branch measurements. From the demagnetized state the applied field is increased from zero to a value ($H_1$), next the applied magnetic field ($H_1$) is decreased to zero, so the magnetic state of the nanomagnets is set at the remanent state $M_z(R1)$, next the ratchet effect is measured. Next magnetic field is applied to the sample up to a value $H_2 > H_1$, and it is decreased back to zero, setting a new remanent state $M_z(R2)$, and so on up to the saturated state is reached, once this state is achieved the maximum value of the ratchet effect remains constant. Three states can be set: $+1$ which corresponds to $M = +M_z$, (in this case $+M_z$ is the saturation magnetization of the hysteresis loop positive branch); 0 which corresponds to $M = 0$; and $-1$ which corresponds to $M = -M_z$, (in this case $-M_z$ is the saturation magnetization of the hysteresis loop negative branch). The zero state is well defined, since a central plateau is measured for low remanent magnetization around the demagnetized state.

**Reading mode (H ≠ 0).** In the second operational mode the nanodevice works as a sensor for magnetic fields. Figure 3 shows the ratchet voltages when a magnetic field is applied parallel to the nanomagnet magnetization direction. Increasing the magnetic field the ratchet signal weakens and a crossover to negative values happens. If the field is further increased dc output voltages enhance again, and the voltage polarity is reversed (Fig. 3(a)). The external fields are applied in the opposite direction to the stray fields in between the triangles, therefore the stray fields are vanishing and the superconductivity is enhanced in these areas[9]. Finally, the applied magnetic field strength is enough to cancel the stray fields, annihilating utterly the vortices. Further increasing the applied magnetic field new vortices emerge with the same polarity that the vortices trapped in the triangles (Fig. 3(b)).

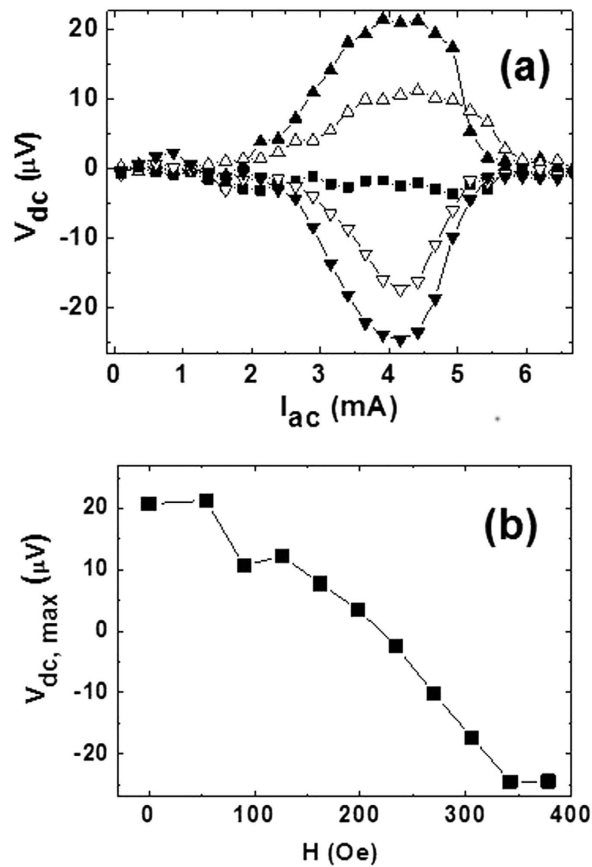

**Figure 3. Ratchet as a function of the applied magnetic field.** (**a**) dc voltages (Y-axis) as a function of the applied ac currents (X-axis) at T = 8.45 K and ac current frequency 10 kHz, for applied magnetic fields: 54 Oe (upward triangles), 126 Oe (upward hollow triangles), 240 Oe (squares), 306 Oe (downward hollow triangles) and 378 Oe (downward triangles). (**b**) maximum output dc voltages (Y-axis) vs. applied magnetic fields (X-axis).

In summary, the nanodevice, placed in an external magnetic field, is very sensitive to the value of the magnetic field and interestingly the output dc voltages comprise from positive to negative, regardless of the magnetic fields do not change polarity.

## Discussion

To figure out the mechanisms which govern these behaviors, we have to considered vortex dynamics. The dynamics of vortex lattice in periodic potential traps have been studied some time ago[14–17]. Many interesting effects have emerged[18], the most relevant effects in connection with the present work are: i) Commensurability effects occur between the vortex lattice and the unit cell of pinning centers; ii) At matching conditions, the vortex lattice motion is slowed down and a minimum appears in the magnetoresistance. Two neighbor minima are always separated by the same magnetic field value;; iii) Different types of vortices participate in these phenomena: a) pinned vortices placed at the pinning centers, b) interstitial vortices[19] or antivortices[20] placed among pinning centers.

To get a better insight into the vortex dynamics of a sample with pinning potentials which are nanomagnets with out of plane magnetization we have fabricated a reference sample with Nb film grown on top of a square array (400 nm × 400 nm) of Cu dots (200 nm diameter and 40 nm thickness), and then we can compare straightforward the well-known behavior of a nonmagnetic sample with the behavior of a magnetic sample with out of plane magnetization. Figure 4 shows the magnetoresistance in the Nb/Cu hybrid film (Fig. 4(a)) and the magnetoresistance data in our nanodevice (Fig. 4(b)). Figure 4(a) shows the minima in the magnetoresistance and as usual, the background resistivity increases when the applied magnetic field is increased. Both effects have been reported in the literature[16]. In the device the magnetoresistance behavior looks very different, see Fig. 4(b).

The background resistance strongly deviates from the usual behavior observed in the reference sample. The interplay between superconducting and magnetism[9] induces a strong enhancement of superconductivity. The superconducting transition temperature is 8.42 K at H = 0 Oe and applying a magnetic field of H = 240 Oe the superconducting transition temperature is 8.50 K (see Fig. 4(b)). The second important experimental feature is that evenly spaced minima arise, being ΔH = 36 Oe the separation between two successive minima, which is similar to the value obtained from the density of pinning centers of the array (see supplementary information). But the most remarkable fact is the value of the dissipation at the minimum resistivity (maximum superconductivity), that corresponds to applied magnetic field around 240 Oe, (dashed line in Fig. 4(b)), the same magnetic field where the crossover occurs (Fig. 3b). Increasing the applied magnetic field from zero to 400 Oe, we begin moving a lattice of interstitial antivortices (placed in between the triangles) which are vanishing when the magnetic field is increasing. At the compensation field (240 Oe) the antivortex lattice is annihilated and a crossover happens. Increasing further the magnetic fields we begin to create vortices which move in the same potential that the antivortices did. Since $\mathbf{E} = \mathbf{B} \times \mathbf{v}$, although the sign of the velocity of both moving lattices is the same, the change of sign of **B** yields a change in the output voltages polarity.

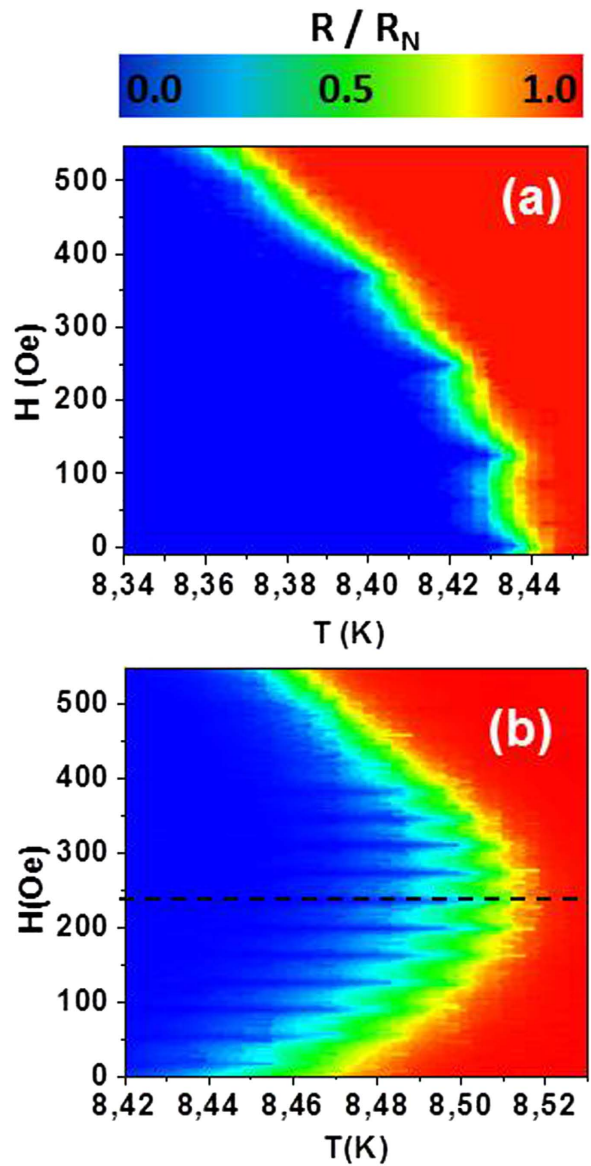

**Figure 4.** Normalized resistance (see color bar) as a function of T (X-axis) and H (Y-axis), $R_N$ being the resistance taken in the normal state (10 K) and measured with 5 μA for: (**a**) Reference sample: Nb film on top of array of Cu nanodots and (**b**) Nanodevice: Nb film on top of array of Co/Pd triangles. The dashed line shows the compensation field position.

In summary, we have fabricated a simple and reliable nanodevice based on the ratchet effect. The interplay between triangular-shaped nanomagnets with out of plane magnetization and superconducting vortices governs the behavior of the device. This nanodevice can operate in two modes depending on whether or not a magnetic field is applied. The first mode (H = 0) is a memory device which can be set in three stable states +1, 0,−1. Stability, robustness and simple readout make this system a good option for a three-state memory device. The same device can work in a second mode (H ≠ 0). In this mode the device is sensitive to small magnetic fields. The main characteristic of this second mode is that the output signal is null for a precise value of the applied magnetic field that only depends on the fabrication characteristics of the nanodevice.

## Methods

Device and sample fabrication: The triangular Co/Pd multilayer and the Cu dot arrays were patterned using standard Electron Beam Lithography (using PMMA resist) on a silicon (100) substrate. Triangle dimensions are 600 nm base and 500 nm height, and they are arranged in a rectangular array with sides 800 nm × 700 nm. After the lithography, a 40 nm thick Co/Pd multilayer with a 3 nm Pd capping was sputtered on top, followed by a lift off process. The Co/Pd multilayers were obtained by magnetron sputtering using two targets and rotatable sample holder, the layer thickness was Co (0.4 nm)/Pd (0.6 nm) and during deposition Ar pressure was 12 mTorr (base pressure was $5 \, 10^{-8}$ Torr). In the case of the array of Cu dots the dimension of the array was 400 nm × 400 nm, and the Cu dot dimensions were thickness 40 nm and diameter 200 nm. On top of these arrays, a 100 nm Nb film was grown using magnetron sputtering, with a base pressure of $5 \, 10^{-8}$ Torr. For electric transport measurements, we defined an 8-terminal cross-shaped bridge using Optical Lithography and Ar/SF6 (1:2) Reactive Ion Etching.

Magnetic characterization: Squid Magnetometer (Quantum Design Magnetic Properties Measurement System) was used to obtain the hysteresis loops as well as the remanent magnetization of both the Co/Pd film and the array of Co/Pd triangles (see Supplementary Information).

Magnetotransport measurements: ac and dc transport properties have been measured in a commercial He cryostat with superconducting solenoid and variable temperature insert (VTI). The temperature controller was Lakeshore 350 and the transport data were taken using an ac/dc current source (Keithley 6221) in combination with a dc nanovoltmeter (Keithley 2182A).

## Acknowledgements

We thank Spanish MINECO grant FIS2013-45469 and CM grant S2013/MIT-2850 and EU COST Action MP-1201. D.G. acknowledges RYC-2012-09864.

## Author Contributions

A.G., E.M.G., J.V. and J.L.V. planned the experiments; J.V., A.G., M.R.O. and D.G. fabricated the samples; J.V. and A.G. carried out the whole experiments; J.V. carried out the magnetic characterization, J.V., A.G. and E.M.G. analyzed the data; J.L.V. wrote the manuscript, E.M.G. and J.L.V. coordinated the whole project. All the authors discussed the results and the manuscript.

## Additional Information

**Supplementary information** accompanies this paper at http://www.nature.com/srep

**Competing financial interests:** The authors declare no competing financial interests.

**How to cite this article**: del Valle, J. *et al.* Superconducting/magnetic Three-state Nanodevice for Memory and Reading Applications. *Sci. Rep.* **5**, 15210; doi: 10.1038/srep15210 (2015).